# Novel non-magnetic hard boride $Co_5B_{16}$ synthesized under high pressure

Elena Bykova,[1,2, a)] Alexander A. Tsirlin,[3] Huiyang Gou,[1,2] Leonid Dubrovinsky,[1] Natalia Dubrovinskaia[2]

[1] *Bayerisches Geoinstitut, Universität Bayreuth, D-95440 Bayreuth, Germany*

[2] *Material Physics and Technology at Extreme Conditions, Laboratory of Crystallography, University of Bayreuth, D-95440 Bayreuth, Germany*

[3] *National Institute of Chemical Physics and Biophysics, Akadeemia tee 23, EE-12618 Tallinn, Estonia*

a) Electronic mail: Elena.Bykova@uni-bayreuth.de

**A first cobalt boride with the Co:B ratio below 1:1, $Co_5B_{16}$, was synthesized under high-pressure high-temperature conditions. It has a unique orthorhombic structure (space group *Pmma*, $a = 19.1736(12)$, $b = 2.9329(1)$, and $c = 5.4886(2)$ Å, $R_1$ (all data) = 0.037). The material is hard, paramagnetic, with a weak temperature dependence of magnetic susceptibility.**

1. Introduction

Transition-metal (TM) borides are interesting in both fundamental and applied aspects. Their high hardness related to the rigid boron network can be superimposed on interesting magnetic and electronic properties driven by the transition-metal ion. For example, $FeB_4$ is a non-magnetic iron boride that becomes superconducting below 2.9 K.[1,2] It is a unique material that combines superhardness and superconductivity.[2] However, it is metastable and can be prepared under high pressure only. In contrast, Fe-rich borides, such as $Fe_2B$ and $FeB$, can be synthesized at ambient pressure. They are ferromagnets with remarkably high Curie temperatures ($T_C$) of 1015 K and 593 K, respectively.[3] On general grounds, one expects that the decrease in the Metal:B ratio will suppress the magnetism,[4] while keeping the system metallic and giving rise to interesting low-temperature effects, such as superconductivity. Therefore, B-rich transition-metal borides remain tantalizing, but also difficult to synthesize.

Cobalt borides share many similarities with the Fe-B compounds. $Co_2B$ and $CoB$ are isostructural to $Fe_2B$ and $FeB$, respectively, but they show a somewhat weaker magnetism. $Co_2B$ becomes ferromagnetic below $T_C$ = 433 K, whereas $CoB$ is a paramagnetic metal.[3] Remarkably, no cobalt borides with the Co:B ratio below 1:1 have been reported. Here, we present synthesis, crystal structure, magnetic properties, and electronic structure of a novel hard boride $Co_5B_{16}$ that fills this gap. This new compound reveals paramagnetic behavior related to a nearly complete occupation of the localized Co $3d$ states.

## 2. Material and methods

### 2.1. Starting materials and synthesis

Single-crystals of $Co_5B_{16}$ were synthesized at pressure of 15 GPa and a temperature of 1873 - 1573 K (heating duration was 40 min) in the Kawai-type multi-anvil apparatus[5] using a 1000-ton (Hymag) hydraulic press. As starting materials we used a cobalt wire (Goodfellow, 99.5% purity) and a boron powder (Chempur Inc., 99.99% purity) which were enclosed into a $h$-BN capsule. The pressure was calibrated based on the phase transitions of standard materials and the temperature was determined using a W3Re/W25Re thermocouple.

### 2.2. Single crystal XRD

A black lustrous prismatic crystal of $Co_5B_{16}$ with a size of 0.07x0.05x0.05 $mm^3$ was used for the crystal structure investigation by means of single-crystal X-ray diffraction. The diffraction data were collected at ambient temperature using a four-circle Oxford Diffraction Xcalibur diffractometer ($\lambda$ = 0.7107 Å) equipped with an Xcalibur Sapphire2 CCD detector. The intensities of the reflections were measured by step scans in omega-scanning with a narrow step width of 0.5°. The data collection and their further integration were performed with the CrysAlisPro software.[6] Absorption corrections were applied empirically by the Scale3 Abspack program implemented in CrysAlisPro. The structure was solved by the direct method and refined by the full matrix least-squares in the anisotropic approximation for all atoms using SHELXTL software[7] implemented in the X-Seed program package.[8] The X-ray diffraction experimental details and crystallographic characteristics of $Co_5B_{16}$ are presented in Table 1 and Table 2. The DIAMOND software[9] was used to create molecular graphics.

The crystallographic data of $Co_5B_{16}$ and further details of the crystal structure investigation have been deposited in the Inorganic Crystal Structure Database[10] and may be obtained free of charge from Fachinformationszentrum Karlsruhe, 76344 Eggenstein-Leopoldshafen, Germany (fax: (+49)7247-808-666; e-mail: crysdata@fiz-karlsruhe.de, http://www.fiz-karlsruhe.de/request_for_deposited_data.html) on quoting the deposition number CSD-427205.

*2.3. Hardness measurements*

Vickers hardness ($H_v$) was measured using a microhardness tester (M-400-G2, LECO Corporation) under loads of 0.5 kgf (4.9 N), 1 kgf (9.8 N) and 1.5 kgf (14.7 N). The average value of hardness was found to be $H_v = 30.1\pm2$ GPa.

*2.4. Magnetic properties*

Magnetic susceptibility was measured with the MPMS SQUID magnetometer in the temperature range 2–380 K in magnetic fields up to 5 T. Heat capacity measurements were attempted with Quantum Design PPMS in zero field using relaxation technique, but no detectable signal could be obtained because of the diminutively small sample size.

*2.5. Electronic structure calculations*

Electronic structure of $Co_5B_{16}$ was calculated in the framework of density functional theory (DFT) using the FPLO code[11] and Perdew-Wang flavor of exchange-correlation potential (LDA).[12] Reciprocal space was sampled with 135 *k*-points in the symmetry-irreducible part of the first Brillouin zone, and the convergence with respect to the number of *k*-points has been carefully checked.

## 3. Results and discussion

*3.1. Crystal structure of $Co_5B_{16}$*

Single-crystals of $Co_5B_{16}$ were synthesized at the pressure of 15 GPa and the temperature of 1873–1573 K. The structure of $Co_5B_{16}$ is orthorhombic (*Pmma* space group, Tables 1-4). Similar to structures of other boron-rich metal borides, it can be described based on a rigid network of boron atoms. In $Co_5B_{16}$ one can easily see honeycomb-like stripes (Fig. 1) oriented along the *b*-axis and condensed into a complicated ramous structure. Such an arrangement of boron atoms gives rise to the straight, channel-like voids along the *b*-axis. Cobalt atoms occupy these voids creating infinite rows. Metal-metal distances in the rows are all equal, but they are larger than the sum of metallic radii of Co atoms (see Table 4). This is similar to the arrangement of metal atoms in other B-rich transition-metal borides, such as $CrB_4$ and $FeB_4$,[2,13] but different from that in $MnB_4$. Although $MnB_4$ has the structure closely related to that of $CrB_4$ and $FeB_4$, Mn–Mn distances in $MnB_4$ are not equal due to the Peierls distortion.[14,15]

Despite some allusion to the tetraboride $CrB_4$ and $FeB_4$ structures, the Co polyhedra in $Co_5B_{16}$ are distinctly different. The Co atoms occupy three independent crystallographic sites, Co(1), Co(2) and Co(3) (Table 2). The structure of $Co_5B_{16}$ can be visualized in terms of packing of three kinds of Co-B polyhedra (Figure 1). An asymmetric part of the structure (Fig. 1*a*) consists of three units: an irregular Co(3)B12 polyhedron, its distorted counterpart Co(1)B12, and a Co(2)B9 polyhedron. Polyhedra of each kind (Co(2)B9, Co(1)B12 and Co(3)B12) pack in columns by sharing common upper and bottom faces and create their own infinite columns parallel to the *b*-axis (Fig. 1*b*). Co(1)B12 polyhedra, as well as Co(3)B12 ones, share the opposite parallelogram-shaped faces, which are parallel to the *ac*-plane. The Co atoms in these columns have the same *y*-coordinates. Co(2)B9 polyhedra pack *via* common triangular faces and each polyhedron sticks to the neighboring Co(1)B12 one through a side quadrilateral face (Fig. 1*b*). As a result Co(1)- and Co(2)- atoms in neighboring columns are shifted on ***b***/2 along the *b*-axis. A polyhedron topologically similar to Co(2)B9 can be deduced from the Co(3)B12 one by removing at once vertices of the two parallelogram-shaped faces of CoB12 and one vertex from the rectangular in the equatorial plain of the latter.

The Co-B distances in Co(3)B12 vary from 2.015(5) to 2.304(4) Å and an average value is 2.183(5) Å (Table 3). Co(1)B12 shares two of its side quadrilateral faces with the Co(2)B9 polyhedra (see Fig. 1). This leads to a distortion of the Co(1)B12 geometry compared to that of Co(3)B12: the Co-B distances' range is 1.991(5) to 2.475(4) Å and an average value increases to 2.275(5) Å. Due to the smaller coordination number of Co(2), the Co(2)B9 polyhedron is the most compact with the average <Co(2)–B> distance of 2.141(7) Å.

Figure 2 provides a comparison of the structure of $Co_5B_{16}$ with that of $MB_4$ tetraborides, where M = Cr, Fe, Mn.[2,13,14] In tetraborides, there is only one kind of $MB12$ polyhedra packed in columns (Fig. 2*b*), so that each column is shifted on ***c***/2 along the *c*-axis with respect to its four nearest neighbors (shown in different colors, light and dark). In $Co_5B_{16}$, every column of Co(3)B12 polyhedra has four neighboring Co(1)B12 columns and shares common B(5) vertices with two of them, while the other two are attached by common edges, which form …–B(6)–B(9)–B(6)–… zigzag chains ( Fig. 2*a*).

The B–B distances in the structure of $Co_5B_{16}$ vary from 1.654(7) to 1.908(7) Å (Table 4). The shortest bond located at the *ac* plane is observed between B atoms of the neighboring Co(3)B12 and Co(1)B12 polyhedra. This is the smallest value of the B-B bond length among transition metal borides with related structures (Table 4). Dense atomic packing and short B-B contacts make $Co_5B_{16}$ rather hard with the measured Vickers hardness $H_v$ = 30±2 GPa, the value slightly higher than reported for $CrB_4$,[13] but lower than that of superhard $FeB_4$.[2]

### 3.2. *Magnetic and electronic properties of $Co_5B_{16}$*

Similar to $FeB_4$, the preparation of single-phase samples of $Co_5B_{16}$ is exceedingly difficult. The largest phase-pure sample available so far is about 0.4 mg and can be used for magnetization measurements only. Magnetic susceptibility shown in Fig. 3 exhibits a weak temperature dependence and a more pronounced field dependence that is likely related to trace amounts of a ferromagnetic impurity. In higher magnetic fields, the impurity signal is suppressed, and a residual temperature-independent susceptibility of about $\chi_0 = 2\times10^{-4}$ emu/mol is observed (Fig. 3). Therefore, $Co_5B_{16}$ behaves as a standard Pauli paramagnet. Small humps in the susceptibility below 100 K require further investigation. Our measurements in low magnetic fields did not show any signatures of superconductivity above 2 K.

Electronic structure of $Co_5B_{16}$ suggests metallic behavior (Fig. 4), with a relatively high density of states at the Fermi level: $N(E_F)$ ~ 1 eV$^{-1}$/Co, similar to 1 eV$^{-1}$/Fe in $FeB_4$.[1] By correcting our experimental $\chi_0$ for the core diamagnetism $\chi_{dia}$ ~ $-2\times10^{-4}$ emu/mol,[16] we arrive at the Pauli contribution $\chi_{Pauli} = \chi_0 - \chi_{dia}$ ~ $4\times10^{-4}$ emu/mol that is comparable, yet larger than the value of $1.6\times10^{-4}$ emu/mol expected from our calculated $N(E_F)$.

The states at the Fermi level are of mixed Co 3*d* and B 2*p* origin, but most of the Co 3*d* states are below the Fermi level and form a relatively narrow band complex between –3 eV and the Fermi level (Fig. 4). These narrow bands should host more localized electrons that tend to become magnetic. In $Co_5B_{16}$, the complete filling of these localized states excludes the magnetism. Indeed, spin-polarized calculations for $Co_5B_{16}$ always converge to the non-magnetic solution.

### 4. Conclusions

In conclusion, the novel boron-rich cobalt boride $Co_5B_{16}$ synthesized at high-pressure and high-temperature conditions was found to be non-magnetic that is in line with the trend of the reduced magnetism upon the decrease in the TM:B ratio. Indeed, $Co_3B$ ($T_C$ = 747 K)[17] and $Co_2B$ ($T_C$ = 433 K)[3] are ferromagnetic, whereas CoB[3] and the $Co_5B_{16}$ are non-magnetic metals. Early studies[18] argued that the behavior of TM-rich borides resembles that of pure transition metals, because boron atoms simply change electron concentration in the TM 3$d$ bands. Apparently, this no longer holds for B-rich TM borides, where a large contribution of boron states is present at the Fermi level (see Fig. 4), and Pauli-paramagnetic behavior is observed.

5. Acknowledgements

A.T. was supported by the Mobilitas grant MTT77 from the ESF. We thank Deepa Kasinathan and Valeriy Verchenko for technical assistance in magnetic susceptibility measurements. H.G. gratefully acknowledges financial support of the Alexander von Humboldt Foundation. The work was supported by the German Research Foundation (DFG). N.D. thanks the DFG for financial support through the Heisenberg Program and the DFG research grant (No. DU 954-8/1).

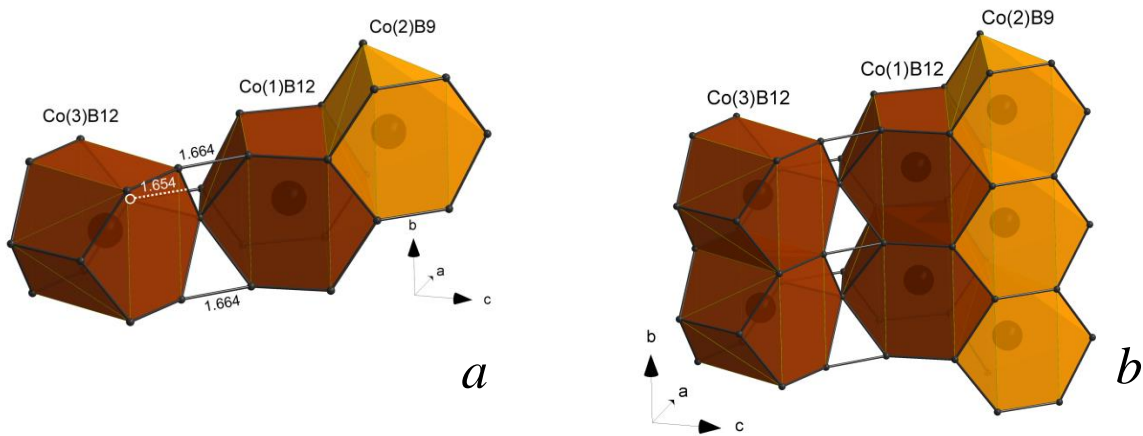

FIG. 1. A polyhedral model of the structure of $Co_5B_{16}$. (*a*) An asymmetric part of the structure consisting of three units: an irregular Co(3)B12 polyhedron, its distorted counterpart Co(1)B12, and a Co(2)B9 polyhedron. (*b*) Packing of the polyhedra in columns along the *b*-axis by sharing common faces. The *y* coordinates of Co atoms in light and dark polyhedra differ by 1/2. B–B bonds are highlighted by bold lines, the shortest distances are labeled.

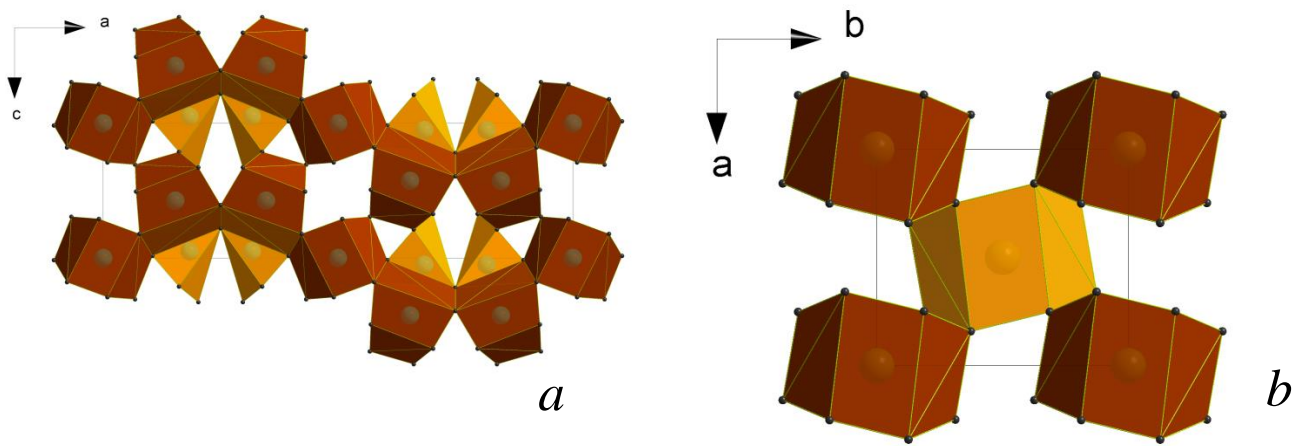

FIG. 2. Comparison of the crystal structures of $Co_5B_{16}$ and $MB_4$ (M = Cr, Fe, Mn)[5,2,6]. (*a*) $Co_5B_{16}$; (*b*) $MB_4$. In the both structures $MB12$ polyhedra pack in columns by sharing common parallelogram-shaped faces either along the *b*- ($Co_5B_{16}$) or *c*-axis ($MB_4$). $Co_5B_{16}$ contains columns constructed of Co(2)B9 polyhedra. Light and dark polyhedra differ in position along *b*- or *c*-axis, respectively.

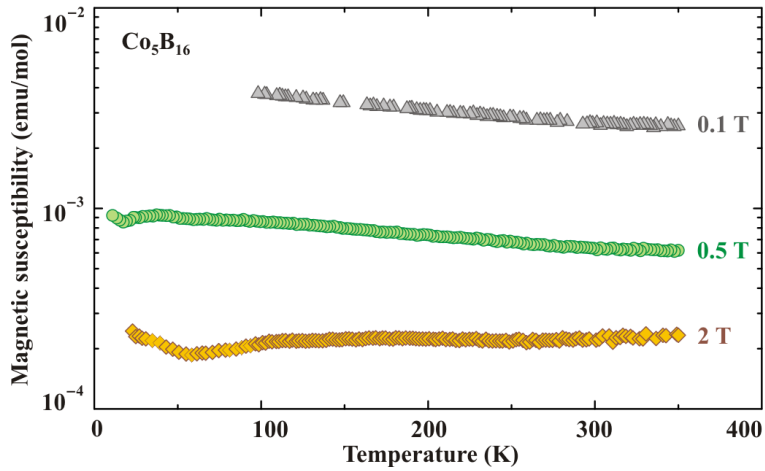

FIG. 3. Magnetic susceptibility of $Co_5B_{16}$ measured in the applied fields of 0.1 T, 0.5 T, and 2 T. In the 0.1 T data, some of the data points were removed because of the low signal and strong noise.

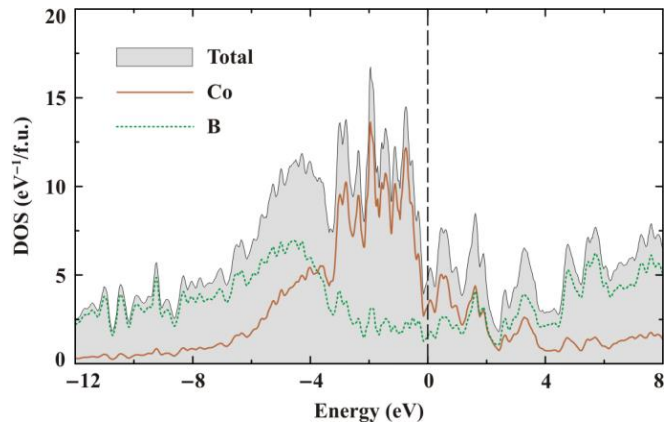

FIG. 4. LDA density of states (DOS) for $Co_5B_{16}$. The total DOS is shown by shading. The solid and dotted lines denote the Co and B contributions, respectively. The Fermi level is at zero energy.

Table 1. Experimental details and crystallographic characteristics for $Co_5B_{16}$

| | |
|---|---|
| Empirical formula | $Co_5B_{16}$ |
| Formula weight (g/mol) | 467.61 |
| Temperature (K) | 296(2) |
| Wavelength (Å) | 0.7107 |
| Crystal system | Orthorhombic |
| Space group | *Pmma* |
| $a$ (Å) | 19.1736(12) |
| $b$ (Å) | 2.93290(10) |
| $c$ (Å) | 5.4886(2) |
| $V$ (Å$^3$) | 308.65(2) |
| Z | 2 |
| Calculated density (g/cm$^3$) | 5.032 |
| Linear absorption coefficient (mm$^{-1}$) | 13.061 |
| F(000) | 430 |
| Crystal size (mm$^3$) | 0.07 x 0.05 x 0.05 |
| Theta range for data collection (deg.) | 3.71 to 30.48 |
| Completeness to theta = 25.00° | 99.7 % |
| Index ranges | $-20 < h < 27$, |
| | $-4 < k < 4$, |
| | $-7 < l < 7$ |
| Reflections collected | 3345 |
| Independent reflections / $R_{int}$ | 569 / 0.0532 |
| Max. and min. transmission | 0.5612 and 0.4617 |
| Refinement method | Full matrix least squares on $F^2$ |
| Data / restraints / parameters | 569 / 0 / 67 |
| Goodness of fit on $F^2$ | 1.145 |
| Final $R$ indices [$I > 2\sigma(I)$] | $R_1 = 0.0282$, $wR_2 = 0.0544$ |
| $R$ indices (all data) | $R_1 = 0.0370$, $wR_2 = 0.0569$ |
| Largest diff. peak and hole (e / Å$^3$) | 0.869 and -0.882 |

Table 2. Atomic coordinates, positions and equivalent isotropic displacement parameters for $Co_5B_{16}$.

| Atom | Wykoff site | x | y | z | $U_{eq}{}^a$, Å$^2$ |
|------|-------------|---|---|---|---------------------|
| Co(1) | 4i | 0.15330(3) | 0 | 0.57345(11) | 0.0048(2) |
| Co(2) | 4j | 0.18369(3) | 0.5 | 0.95008(12) | 0.0046(2) |
| Co(3) | 2a | 0 | 0 | 0 | 0.0045(2) |
| B(1) | 2e | 0.25 | 0 | 0.7796(14) | 0.0067(14) |
| B(2) | 4i | 0.1614(3) | 0 | 0.2119(9) | 0.0044(9) |
| B(3) | 4j | 0.0700(3) | 0.5 | 0.5066(10) | 0.0062(9) |
| B(4) | 2f | 0.25 | 0.5 | 0.6068(14) | 0.0073(14) |
| B(5) | 4i | 0.678(3) | 0 | 0.3065(10) | 0.0061(10) |
| B(6) | 4i | 0.1050(2) | 0 | 0.9840(10) | 0.0048(9) |
| B(7) | 4j | 0.2026(3) | 0.5 | 0.3273(10) | 0.0055(9) |
| B(8) | 4j | 0.0091(3) | 0.5 | 0.2906(9) | 0.0064(10) |
| B(9) | 4j | 0.0772(3) | 0.5 | 0.8208(10) | 0.0056(10) |

$^a$ $U_{eq}$ is defined as one third of the trace of the orthogonalized $U^{ij}$ tensor.

Table 3. Co…B interatomic distances in CoB12 and CoB9 polyhedra (in Å)

| **Co(1)B12** | | | **Co(2)B9** | | | **Co(3)B12** | | |
|---|---|---|---|---|---|---|---|---|
| Co(1)–B(2) | 1x | 1.991(5) | Co(2)–B(2) | 2x | 2.097(4) | Co(3)–B(2) | 2x | 2.015(5) |
| Co(1)–B(1) | 1x | 2.172(4) | Co(2)–B(7) | 1x | 2.102(5) | Co(3)–B(7) | 2x | 2.126(5) |
| Co(1)–B(5) | 1x | 2.198(5) | Co(2)–B(6) | 2x | 2.112(3) | Co(3)–B(6) | 4x | 2.174(4) |
| Co(1)–B(3) | 2x | 2.199(4) | Co(2)–B(1) | 2x | 2.155(3) | Co(3)–B(1) | 4x | 2.304(4) |
| Co(1)–B(7) | 2x | 2.206(4) | Co(2)–B(9) | 1x | 2.162(5) | | | |
| Co(1)–B(4) | 2x | 2.3709(8) | Co(2)–B(4) | 1x | 2.273(7) | | | |
| Co(1)–B(6) | 1x | 2.436(5) | | | | | | |
| Co(1)–B(9) | 2x | 2.475(4) | | | | | | |
| **<Co(1)–B>** | | **2.275(5)** | **<Co(2)–B>** | | **2.141(7)** | **<Co(3)–B>** | | **2.183(5)** |

Table 4. Interatomic distances in metal borides with related structures.

| Metal boride | $M$–$M$ distances, Å | $M$–B distances, Å | B–B distances, Å | Reference |
|---|---|---|---|---|
| $Co_5B_{16}$ | 2.9329(1) | 1.991(5)–2.475(4) | 1.654(7)–1.908(7) | This work |
| $CrB_4$ | 2.8681(5) | 2.053(4) | 1.743(6) | [13] |
| | | 2.153(4) | 1.835(4) | |
| | | 2.178(3) | 1.868(6) | |
| | | 2.261(3) | | |
| $FeB_4$ | 2.9990(3) | 2.009(4) | 1.714(6) | [2] |
| | | 2.109(4) | 1.8443(3) | |
| | | 2.136(3) | 1.894(6) | |
| | | 2.266(3) | | |
| $MnB_4$ | 2.7006(6), 3.1953(7) | 1.999(4)–2.310(4) | 1.703(6)–1.893(8) | [14] |